\documentclass[sigconf]{aamas}

\usepackage{balance}

\makeatletter
\gdef\@copyrightpermission{
  \begin{minipage}{0.2\columnwidth}
   \href{https://creativecommons.org/licenses/by/4.0/}{\includegraphics[width=0.90\textwidth]{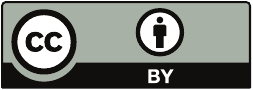}}
  \end{minipage}\hfill
  \begin{minipage}{0.8\columnwidth}
   \href{https://creativecommons.org/licenses/by/4.0/}{This work is licensed under a Creative Commons Attribution International 4.0 License.}
  \end{minipage}
  \vspace{5pt}
}
\makeatother

\setcopyright{rightsretained}
\acmConference[Preprint]{Preprint: submitted to AAMAS 2027 (Blue Sky Ideas Track)}{2026}{arXiv}{Under review}
\copyrightyear{2026}
\acmYear{2026}
\acmDOI{}
\acmISBN{}
\submissionType{Blue Sky Ideas Track}

\usepackage{xspace}
\newcommand{\definedas}{\ensuremath{:=}}

\begin{document}

\title{A Self-Evolving Agent Runs Like a Gas: Describing and Predicting Its Evolution with Fluid Dynamics}

\author{Haipeng Wang}
\affiliation{%
  \institution{Neusoft Corporation}
  \country{China}}
\email{haipeng\_wang@neusoft.com}

\begin{abstract}
A self-evolving LLM agent improves in service by rewriting its own plugin library under a validation gate. Agent engineering is converging on exactly such agents: single systems in continuous operation, carrying populations of runtime instances---plugins, skills, tools---past the point where anyone can read the library and still predict what the system will do. Yet the run itself---the aggregate motion of that population across hundreds of edit rounds---is unstudied: no theory says whether it relaxes, oscillates, or freezes. A recent kinetic theory supplies the missing level: under four architectural properties, the plugin instances of one agent behave as identical hard spheres whose density obeys a master equation---and the theory is candid that its gate injects and drains credit, momentum is not conserved, and waves are excluded by construction. We argue the boundary is architectural; keeping the four safety theorems, we stack conservation axioms on top. A closed-credit ledger makes credit a collision invariant and unlocks acoustic waves at a measurable, engineerable speed; a zero-sum projection gate conserves centroid momentum unconditionally, for any proposal stream. Both layers are proven and executed: the axis total is conserved to the digit over 66 interventions under a live gate, and the pre-registered decisive probe returns the theory's directional verdict---linear wavefronts where the ledger holds, dissipation where it does not---replicated across two population sizes. For agents that must never stop, the gas description is not a metaphor: it is the measurement interface through which their future becomes readable.
\end{abstract}


\maketitle

\section{Introduction}\label{sec:intro}

The destination of current agent engineering is an operating system. Academic prototypes already embed LLM agents into OS-style architectures~\cite{aios2024}; industrial harnesses converge on everything-is-a-plugin, whose engineering point is \emph{hot-swappable capability}: tools mount, unmount, and update while the operation keeps running. The trend has consolidated money: Alibaba Cloud's September 2026 Qwen Book ``agent computer,'' stated principle \emph{OS-as-harness}~\cite{omarchy2026}. The empirical caution agrees: split capability across many cooperating agents and \emph{coordination} becomes the dominant failure mode; multi-agent pipelines frequently lose to a single agent with good tools~\cite{cemri2025mast,anthropic2024agents}. Four lines (academic prototypes, industrial consolidation, failure evidence, engineering practice) converge on one design point: \emph{one agent} in continuous operation, a population of runtime instances growing inside it (MCP servers, plugins, skills, tools, ...). The complexity to master is not a society of agents; it is the population inside one agent. Past a few hundred instances, no one reads the whole library and still understands the system: the run becomes \emph{unpredictable}, because instances interact and compose in combinations nobody can enumerate.

DeepSeek, born inside quantitative trading where the operation cannot pause for a recompile, shipped the DeepSeek Harness (DSH), a discipline where \emph{everything is a plugin}: every runtime instance written, composable, auditable, and swappable~\cite{deepseekharness}. Under that discipline self-evolving agents become trainable, the loop well studied: proposing edits, keeping it safe, pricing acceptances~\cite{luo2026skillopt,sgm2025statistical,shinn2024reflexion,yin2025godel}. What has not been studied is the agent's \emph{run} itself, the aggregate motion of the trainable state across hundreds of edit rounds. Does the running system obey any law: does it relax, oscillate, freeze? Can its endpoint be predicted without running it? The literature has no answer: no level of description at which those questions make sense. The agent as a unit is too coarse: the interesting motion is inside it.

The missing level turned out to be physics. In studying DSH, a recent kinetic theory~\cite{ktp2026} observed that the 2026 Fields Medal work of Deng (deriving the Boltzmann equation from deterministic hard-sphere dynamics, the mathematical heart of Hilbert's sixth problem~\cite{hilbert1902,deng2024longtime}) is exactly the needed move: from microscopic rules to a macroscopic equation of motion. That theory is candid that its fluid layer is locked: no conserved quantity survives its operators, so waves, vortices, and cascades are excluded by construction. We read that diagnosis as a design menu, one constraint up front: the four safety properties are the load-bearing foundation of hot-swap safety, proven theorems, and this program does not touch them. \emph{Phenomena are not seen or missed; they are admitted or excluded by axioms}, so the program adds axioms rather than removing them: stack conservation laws on top of the untouched foundation, and inherit each new layer's fluid regime with its equations, its phenomena, and its instruments.

The program, stated in one sentence: \emph{keep the safety axioms A1--A4 untouched, stack conservation axioms on top of them, derive what fluid phenomena each stacked layer admits, and use the inherited theory of that layer to describe and predict the DSH-plugin agent's run.} The paper proceeds in five steps: the gas the four axioms give and why its fluid layer stays locked (\ref{sec:locked}), the stack we add (\ref{sec:stack}), what it predicts (\ref{sec:phenomena}), the verification (Section~\ref{sec:experiments}), the outlook (Section~\ref{sec:roadmap}). Our contributions:
\begin{itemize}
\item \textbf{A conservation stack for plugin agents.} Two new axioms (a closed-credit ledger and a zero-sum projection gate) layered on the untouched safety foundation: proofs that the ledger makes credit a collision invariant (a barotropic Euler regime, acoustic waves at engineerable speeds), and that the gate conserves centroid momentum \emph{unconditionally}, for any proposal stream.
\item \textbf{Phenomena and instruments with named equations.} A forecastable credit-shock failure mode, collective ordering, and six instruments from statistical physics that read population statistics any runtime logs, composed into a differential diagnosis for stalled training.
\item \textbf{Executed verification.} Both laws execute through a live gate to the digit (66 interventions under free, non-conserving proposals), and the pre-registered decisive probe returns the directional verdict (wavefronts where the ledger holds, dissipation where credit washes), replicated across two population sizes.
\end{itemize}

\section{Related Work}\label{sec:related}

Mean-field and kinetic approaches to multiagent learning fix the architecture and study the dynamics~\cite{pareschi1999,mfmar2018,jung2026}; self-evolution research fixes a loop and studies its safety or gains~\cite{luo2026skillopt,sgm2025statistical}. This program inverts both: the \emph{architecture is the free variable that selects the class of dynamics}, the axioms its coordinate system. And where multi-agent designs disperse capability, inheriting their coordination failures~\cite{cemri2025mast}, this program concentrates it inside one agent: the gas of Section~\ref{sec:locked} is one agent's interior. Agent-OS architectures supply the habitat~\cite{aios2024,omarchy2026}; the kinetic ladder is classical mathematics~\cite{deng2024longtime,golse2004ns,haff1983,bressan2000}; manufactured emergence~\cite{reynolds1987,gardner1970,acebron2005} yields patterns without equations; this program's lineage laws come with measurable parameters. What is new is the substrate: conservation laws as \emph{legislated runtime policy} rather than physical law. Axioms, not scale, are the levers.

\section{Theory: The Conservation Stack Program}\label{sec:theory}

The program builds on a foundation it did not lay, which is stated first; proofs and per-run artifacts are deferred to the extended version.

\subsection{Step Zero: The Gas the Four Axioms Give}\label{sec:locked}

DSH's Cordis kernel~\cite{shi2026composability} proves four structural properties of plugins: \textbf{A1} permutation symmetry (order-independent composition), \textbf{A2} mount/unmount reversibility (no residue), \textbf{A3} acyclic dependencies (no instantaneous self-influence), and \textbf{A4} typed contracts (bounded, short-range interaction). The kinetic theory~\cite{ktp2026} shows that under exactly these four properties, the plugin-version instances inside one agent behave as a gas of identical hard spheres, writing a master equation over the plugin$\times$version$\times$task grid,\begin{equation}\label{eq:ktp}
\partial_t f + v\!\cdot\!\nabla_x f + \nabla_{\!v}\!\cdot\!(F f) \;=\; Q[f,f] + R[f] + G[f],
\end{equation}
with collision $Q$ (co-activation with attribution), reaction $R$ (mount\slash unmount), gate $G$ (selective dissipation), and external field $F$ (the task distribution). Because the plugin library is the agent's \emph{entire} trainable state, this is not an analogy layered on top of the agent: the agent's run is, at this level of description, this gas's dynamics. Its headline signatures are already measured under pre-registered criteria~\cite{ktp2026}: NESS relaxation (not convergence to an optimum), and $-1/2$ fluctuation scaling at sufficient edit density ($\hat\beta=-0.58$ vs.\ $-1/2$; flat at $-0.04$ when sparse).

The lock is structural: the classical route from kinetics to hydrodynamics runs through \emph{collision invariants}, conserved fields that survive the collision operator and close the moment hierarchy at Euler level~\cite{golse2004ns}. The agent's gas of Eq.~\eqref{eq:ktp} has none: the gate injects and drains credit, reward arrives from outside, and the collision rule mints attribution mass. Every candidate invariant is annihilated at the collision layer.

\begin{corollary}[No transport without conservation]\label{cor:notransport}
Under the current axioms (open credit), for every first-order moment $\varphi$ some operator $T \in \{Q, R, G, F\}$ has $\int \varphi\,(Tf)\neq 0$; hence no conservation law and no closed $p$-system (the classical route to finite-speed transport), so within Eq.~\eqref{eq:ktp} no library moment carries a closed transport equation (the gate adjudicates on scores, not on moments), and open-credit responses must be mean-field or dissipative (Eq.~\eqref{eq:regimes}).
\end{corollary}

\subsection{Step One: The Conservation Stack}\label{sec:stack}

The first rule is restraint: A1--A4 are the kernel-proven foundation of hot-swap safety, kept in force in every runtime below; what the program adds is a second storey, conservation axioms stacked on top: a closed-credit ledger, a zero-sum projection gate, an energy analogue left open. Relaxing A1--A4 is demolition, not extension; our release experiments (Section~\ref{sec:experiments}) quantify the named failure modes: evidence \emph{for} the foundation.

\subsubsection{Layer 1: the closed-credit ledger}\label{sec:conservation}

Call a runtime \emph{closed-credit} if reward is minted once and thereafter only \emph{transferred} among instances, no operator destroys credit, and the gate adjudicates on \emph{shares} of the conserved total: each clause is an engineering decision, and the theorem below is a build sheet.

\begin{proposition}[Conditional Euler regime]\label{prop:euler}
Let $f$ solve Eq.~\eqref{eq:ktp} on a closed-credit runtime, with the bounded-locality collision structure of~\cite{ktp2026}. Then $M \definedas \int e\,f\,dx\,dv\,de$ is a collision invariant; \emph{given the virial constitutive closure below}, the moment hierarchy closes at first order in the conserved field as a hyperbolic $p$-system, and credit-density perturbations propagate with finite speed $c^2 = \partial p/\partial\rho$.
\end{proposition}

The constitutive closure is the surface an architect tunes: attribution transfers of second moment $\kappa_a \definedas \tfrac{1}{2}\int (\Delta e)^2\,\mathrm{d}P(\Delta e)$, played against per-cell co-activation traffic $r_c \propto \rho_c$, amount to the Haff-type equation of state $p = \kappa_a\rho^2$~\cite{haff1983}; the conserved pair $(\rho,\, m)$ with $m \definedas \rho\, u$ obeys the $p$-system $\partial_t \rho + \nabla_{\!x}\!\cdot\!(\rho\, u) = 0$,\; $\partial_t (\rho\, u) + \nabla_{\!x}\!\cdot\!\big(\rho\, u \otimes u\big) + \nabla_{\!x} p(\rho) = \rho\, F$, whose characteristic speeds $u \pm c$ carry the acoustic speed
\begin{equation}\label{eq:speed}
c^2 = p'(\rho) = 2\kappa_a\rho \;>\; 0,
\end{equation}
so hyperbolicity is immediate and the Cauchy problem is well-posed~\cite{bressan2000}. Because the runtime is discrete with a complete ledger, this balance is \emph{exact}: cellwise $E_c(t{+}1)-E_c(t)=\sum_{c'}(\Phi_{c'\to c}-\Phi_{c\to c'})$, no error term, a ledger identity auditable from the training log.

\subsubsection{Layer 2: zero-sum attribution, and its strict form}\label{sec:momentum}

The next rung is momentum: attribution updates are unconstrained, so internal collisions move $\sum_i v_i$ at will. The \emph{zero-sum attribution axiom} legislates Newton's third law: every co-activation's attribution updates sum to zero on a shared momentum axis ($\eta_i a_i + \eta_j a_j = 0$, where $\theta_i \leftarrow \theta_i + \eta_i a_i$ is the position update), with a birth/death protocol so mounting and unmounting mint no position. On such a runtime the axis total $P=\sum_i\theta_i$ is a second collision invariant, provided attribution pairs are bundled into single edit sets so the gate cannot split a zero-sum pair.

Legislation has a weakness: it holds only if proposals comply; gases conserve momentum because the collision operator cannot do otherwise. The strict form is an operator:

\begin{quote}
\textbf{Projection axiom (A5-v, mechanistic form).} The runtime does not trust proposals: every candidate edit set, before evaluation, passes the \emph{zero-sum projection}: its net axis displacement $d$ is computed, and the deficit $-d$ is charged to a designated \emph{reserve instance}. Conservation is a property of the application operator, not of the proposals.
\end{quote}

\begin{proposition}[Unconditional centroid conservation]\label{prop:momfield}
Under the projection gate, for \emph{every} proposal stream, including streams that violate any axiom, the accepted library satisfies $\sum_i \theta_i^{(t+1)} = \sum_i \theta_i^{(t)}$ across every acceptance: the centroid's coordinate on the momentum axis (not a credit total) is conserved unconditionally. Under the locality and virial assumptions of Proposition~\ref{prop:euler}, the joint $(\rho, g)$ first-order system is hyperbolic with two acoustic branches $c_\rho = \sqrt{2\kappa_a\rho}$ and $c_v = \sqrt{2\kappa_v g}$.
\end{proposition}

Concretely, the gate applies $\mathcal{P}(E) = E \cup \{(\mathrm{res}, -d(E))\}$ when the reserve charge is admissible, else drops the set; either way $d(\mathcal{P}(E)) = 0$, and every set the gate evaluates lies in the zero-sum class. The reserve is the wall: it takes the recoil, is named, and is auditable. In one dimension each conserved pair forms a $2\times2$ $p$-system with its own acoustic branch and Riemann invariants, with credit and displacement waves uncoupled at first order. The stack \emph{composes}: a further layer never breaks a conservation law already in force (stack monotonicity), which licenses the word. Layer 3, the energy analogue, is the open rung (Section~\ref{sec:roadmap}).

\subsection{Step Two: Predicted Phenomena and Instruments}\label{sec:phenomena}

``Wave'' is operationalized before it is claimed: a perturbation profile whose radius grows \emph{linearly} in time, as opposed to instant or decaying:
\begin{equation}\label{eq:regimes}
\underbrace{r(t) = c\,(t - t_0)}_{\text{wave-like (finite speed)}},
\quad
\underbrace{r \equiv R,\ \text{instant and global}}_{\text{mean-field}},
\quad
\underbrace{\mathrm{resp}(r) \propto e^{-r/\xi}}_{\text{dissipative}};
\end{equation}
$r \ge 0$ is the response radius (instances responding, zero before onset $t_0$), and the dissipative profile is read on $r \ge 0$;
Corollary~\ref{cor:notransport} confines the locked runtime to the second and third patterns, so the closed-credit prediction is directional; a measurable speed is a settable requirement: damage containment and intervention timing become wave-speed questions.

\textbf{A forecastable failure mode: the credit shock.} A $p$-system's classical theory is not gentle: compression fronts steepen and shocks form in finite time, a gradient blowup kin to the announced Navier--Stokes construction~\cite{openai2026ns}. Translated: if credit density somewhere grows faster than the acoustic flux can carry it away, the stack's own equation predicts a discontinuity: a \emph{credit shock}, the avalanche-like concentration practitioner folklore calls \emph{collapse} or \emph{winner-takes-all}. We do not claim observability at minimal scale; the theory names the regime (compression against a bounded acoustic flux) and the quantity to watch ($\lvert\nabla(\rho u)\rvert$ against $c_\rho$), a \emph{forecastable} failure mode before it is ever observed.

\textbf{Ordering, and the instruments.} A second predicted collective is synchrony: Kuramoto-type populations snap from disorder to order at a critical coupling, a jump invisible in any individual's telemetry, with the order billed to the environment~\cite{prigogine1978}; the correlation length $\xi$ (diverging) and the four-point susceptibility $\chi_4$ (peaking) are the early-warning gauges. Synchrony sits outside the stack (locking needs dependency cycles, the deliberately unsafe family), but its moral is the paper's: the collective is measurable where the individual is not. Because conservation laws fix what the macroscopic variables \emph{are}, the measurement technology is a century old: six instruments read any runtime's population logs (aging $\tau$, correlation length $\xi$, effective temperature $T_{\mathrm{eff}}$, transport coefficients $D,\eta$, dynamical heterogeneity $\chi_4$, and fluctuation scaling, field-tested~\cite{ktp2026}). Table~\ref{tab:diagnosis} composes them into a differential diagnosis: the three causes of a stalled run separate by joint readings. The Blue Sky claim: debugging a stalled agent should look like a lab session: perturb, relax, measure, rather than read telemetry and guess.

\begin{table*}[t]
\centering
\footnotesize
\caption{Differential diagnosis for stalled training: three plateau causes separate by joint readings; prescriptions differ.}
\label{tab:diagnosis}
\begin{tabular}{@{}p{3.9cm}p{2.0cm}p{3.3cm}p{1.8cm}p{5.0cm}@{}}
\toprule
Reading & Aging $\alpha$ & Gate info ($\delta_{\mathrm{gate}}$ vs $\delta_{\mathrm{ind}}$) & Accept rate & Cause $\to$ prescription \\
\midrule
$\tau \sim t_w^{\alpha}$, $\alpha > 0$; $\chi_4$ high & $>0$ & both $\approx 0$ & varied & structural freezing $\to$ anneal \\
$\tau$ flat; fresh tasks restore gains & $0$ & $\delta_{\mathrm{gate}}>0,\ \delta_{\mathrm{ind}}\approx 0$ & normal & information exhaustion $\to$ refresh validation \\
$\tau$ flat; gains stay absent everywhere & $0$ & both $\approx 0$ & $\to 0$ & proposal exhaustion $\to$ enrich reflector / raise edit budget \\
\bottomrule
\end{tabular}
\end{table*}

\section{Experiments}\label{sec:experiments}

All experiments run on the faithful minimal instance (WebShop text mode~\cite{yao2022webshop}; 50--200 four-parameter skill plugins; deterministic stub reflector; strict validation gate), protocols frozen before the runs, numbers reproducible from the archived run artifacts.

\textbf{Foundation stress-tests.} Releasing A4 (typed contracts) lets one out-of-domain edit poison the highest-value document: the improvement waterline stops dead ($0.2158$ vs.\ $0.2683$ matched control). Releasing A3 (acyclicity) lets dependency cycles form between the two highest-value documents: $63$ execution aborts vs.\ zero in control, mean score \emph{untouched}, since top-3 retrieval routes around the dead document. Both bites are the theory's named failure channels; the second is the strongest practical argument for instruments: score-class readings are blind to structure-class failures.

\textbf{Channel calibration.} Perturbing the highest-retrieval document against an undisturbed twin: execution-channel disturbances do not propagate at all (exactly zero response, 6 of 6: the channels are decoupled, so damage can be invisible to retrieval telemetry); retrieval-channel disturbances propagate mean-field (full-library response within one round); the conservation channel (moving credit itself) is the one that can carry a wave.

\textbf{Conservation executed.} The momentum layer promises conservation through a live gate for \emph{any} proposal stream. Legislated (bundled zero-sum proposals): the axis total fixed to the digit in every seed. Mechanistic (free non-zero-sum proposals through the projection gate): the projection fired 66 times, the axis total conserved exactly in every seed, while the unprojected control drifted, at no measurable performance cost. Identical free streams; only the projected arm conserves: Proposition~\ref{prop:momfield} verified. Still unobserved: the displacement wave itself (the $c_v$ branch), which needs the wave protocol on a projected, closed-credit arm.

\textbf{The decisive probe.}\label{sec:probe} The theory's directional prediction (waves only where the ledger holds) was pre-registered with frozen criteria (Eq.~\eqref{eq:regimes}), then executed: closed-credit vs.\ an open control whose gate re-allocates $15\%$ of total credit per round, twins forked in-process from the identical pre-disturbance state, conservation audited to machine precision every round. Three findings. \emph{First, the directional prediction holds}: the closed arm shows linear wavefronts in 2 of 5 seeds ($R^2=0.94$/$0.91$ over 20/16 rounds) at $0.41$ and $0.65$ instances per round, the first reading of $c_\rho$'s discrete counterpart, while the open control shows none (5 of 5 dissipative, exactly Corollary~\ref{cor:notransport}'s verdict). \emph{Second}, mean-field is absent here for both economies (0 of 10): the channel taxonomy is complete. \emph{Third}, the absolute criterion (3 of 5) was not met (2 of 5), and we report that as it stands.

\begin{figure}[t]
\centering
\includegraphics[width=0.74\columnwidth]{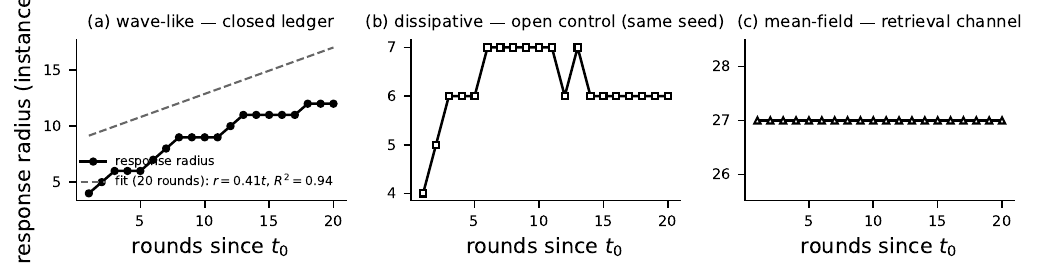}
\caption{(a)~Closed-credit arm, seed 31: the response radius (instances with credit deviation $>0.002$ from the in-process twin, perturbed pair excluded) grows linearly (frozen-criteria fit of Table~\ref{tab:cwave}: $R^2=0.94$, $0.41$ instances/round over 20 rounds). (b)~The open-credit control, same seed: saturates without a sustained wavefront. (c)~Retrieval-channel disturbance: full-neighborhood response within one round (mean-field).}
\label{fig:modes}
\end{figure}

\begin{table}[t]
\centering
\footnotesize
\caption{The decisive probe (five seeds per arm, twin-controlled, frozen criteria). ``Wave-like'' = linear radius growth ($R^2 \ge 0.9$, $\ge 5$ rounds); the wave seeds ran linear for 20 and 16 rounds at $0.41$/$0.65$ instances/round.}
\label{tab:cwave}
\begin{tabular}{@{}llccl@{}}
\toprule
Arm & seed & verdict & best $R^2$ & max radius/48 \\
\midrule
closed & 11 & dissipative & --- & 2 \\
closed & 21 & dissipative & --- & 12 \\
closed & 31 & \textbf{wave-like} & 0.94 & 12 \\
closed & 41 & dissipative & --- & 13 \\
closed & 51 & \textbf{wave-like} & 0.91 & 14 \\
\midrule
\multicolumn{5}{@{}l@{}}{open (5 seeds): all dissipative, $R^2$ n/a, max radius 7--11} \\
\bottomrule
\end{tabular}
\end{table}

\textbf{Larger population ($N{=}200$).} Re-run at four times the library (10 seeds per arm, one lost to a logging collision; component-aware perturbation: the co-activation graph's largest component holds 93 of 200): the verdict replicates, 2 of 9 closed vs.\ 0 of 10 open; mean-field again absent (0 of 19); normalized reach larger on the closed arm ($0.71$ vs.\ $0.60$). The batch also corrects the small-instance reading: the binding constraint is \emph{signal attenuation}, because rewards split over four times the accounts dilute the disturbance within a few hops, so scaling propagation is a \emph{ledger-density} lever, not a theory limitation. Two pre-registered side measurements returned null: the $\sqrt{\varphi}$ speed scaling fell below the noise floor; relaxation time was flat in waiting time, the driven steady state, not aging. Scope: one task domain, a stub reflector; qualitative signatures carry by construction, constants are device-local; a richer reflector could reach regimes the stub cannot.

\section{Outlook and Open Problems}\label{sec:roadmap}

\textbf{The blue sky.} Take self-evolving agents' trajectory seriously: recursive self-improvement over a plugin library that is the entire trainable state~\cite{deepseekharness}. Its endpoint: an operating-system-like agent, millions of versioned plugins trading credit under gates. At those scales this paper's questions become operating requirements: \emph{how fast does damage travel? is the population aging? which conservation laws held today?} And a harder constraint: a downtime-intolerant agent (the trading floor, where the book cannot pause to be debugged) admits no offline inspection; its future must be read off the live run: the gas description's readouts, as statistical mechanics made Avogadro-scale populations governable. For agent populations that science does not exist yet; this paper builds its first two rungs. For the practitioner this is already an operations manual: install the rules, read the dashboard, diagnose the stall (Table~\ref{tab:diagnosis}), watch the collapse gauge; a one-page quick-start is in the extended version.

\textbf{Open problems.} (1)~\emph{Axiom independence}: is each of A1--A4 independent in the kernel's theorem graph~\cite{shi2026composability}? (2)~\emph{The energy analogue}: which plugin quantity plays energy (Noether-style: pair kernel symmetries with candidates)? (3)~\emph{Architecture contracts}: specify a blast radius as a wave-speed requirement. (4)~\emph{Heterogeneous populations}: mixed-gas physics for mixed axiom sheets, and instruments as standards.

\section{Conclusion}\label{sec:conclusion}

Four structural axioms turn the agent's interior into a gas: signatures measurable, failure channels nameable, a conservation stack unlocking a fluid regime whose directional prediction replicates across two scales. The gas description is not a metaphor laid over the agent; it is a measurement interface for it.

\balance
\bibliographystyle{ACM-Reference-Format}
\bibliography{references}

\end{document}